\input epsf
\input harvmac
%
  \def\CC{{\cal C}} \def\CD{{\cal D}}
   \def\CH{{\cal H}}
   \def\CL{{\cal L}}
\def\CM{{\cal M}}  \def\CO{{\cal O}} \def\CP{{\cal P}}
  \def\CS{{\cal S}} \def\CT{{\cal T}}
   
 \def\CZ{{\cal Z}}

\font\fivebf=cmbx5
\def\rvec{{\bf \vec r}} 
\def\rvecR{{\bf \vec r}_{\hbox{\fivebf R}}} 
\def\bR{b_{\hbox{\fivebf R}}}
\def\kvec{{\bf \vec k}}

\def\qvec{{\bf \vec q}}
\def\zerovec{{\bf \vec 0}}

\font\twelverm=cmr10 scaled \magstep1
\def\npr{\hbox{\twelverm :}}
\def\ii{{\rm i}}
\def\sprod{\mathop{\Pi}}
\def\sprodp{\mathop{\Pi'}}
\def\ssum{\mathop{\Sigma}}
\def\sssum{\mathop{{\scriptstyle\Sigma}}}
\def\blangle{\bigl\langle}
\def\brangle{\bigr\rangle}

\def\RR{\relax{\rm I\kern-.18em R}}
\def\Rr{\relax{\ninerm I\kern-.22em \ninerm R}}
\def\setminusp{\hbox{$/ \kern -3pt {}_p$}}
\def\ssetminusp{\hbox{$\scriptstyle / \kern -2pt {}_p$}}
\def\lesssim{\hbox{\raise.4ex \hbox{$<$} \kern-1.1em \lower.8ex \hbox{$\sim$}}}

\Title{SPhT/93-074}{
\vbox{
\vskip -2truecm
\centerline{Renormalization and Hyperscaling for} 
\centerline{Self-Avoiding Manifold Models}   
}}
\centerline{Fran\c cois David\footnote{$^\dagger$}
{Member of CNRS},
Bertrand Duplantier{$^\dagger$}{} and Emmanuel Guitter}
\bigskip\centerline{Service de Physique Th\'eorique\footnote{$^\star$}{
Laboratoire de la Direction des Sciences de la Mati\`ere du Commissariat
\`a l'Energie Atomique}}
\centerline{C.E. Saclay}
\centerline{F-91191 Gif-sur-Yvette, France}
\vskip 1.truecm
\centerline{\bf Abstract}{
\ninerm
\textfont0=\ninerm
\font\ninemit=cmmi9
\font\sevenmit=cmmi7
\textfont1=\ninemit
\scriptfont0=\sevenrm
\scriptfont1=\sevenmit
\baselineskip=11pt 
\bigskip
The renormalizability of the self-avoiding manifold (SAM) Edwards model is
established.
We use a new short distance multilocal operator product expansion,
which extends methods of local field theories to a large class of models with
non-local singular interactions.
This validates the direct renormalization method introduced before, as well as
scaling laws.
A new general hyperscaling relation is derived.
Manifolds at the $\Theta$-point and long range Coulomb interactions are briefly
discussed.
\bigskip
\par
}
\vskip .3in
\Date{PACS numbers: 05.20.-y, 11.10.Gh, 11.17.+y}
\hfuzz 1.pt
\vfill\eject

The statistical mechanics of fluctuating surfaces has attracted
much attention in the recent years, with applications in many areas of
physics, from string theories in high energy physics to interface and membrane
problems in soft condensed matter physics and biophysics
\ref\rJerus{{\sl Statistical Mechanics of Membranes and
Surfaces}, Proceedings of the Fifth Jerusalem Winter School for Theoretical
Physics (1987), D. R. Nelson, T. Piran and S. Weinberg Eds., World Scientific,
Singapore (1989).}.
In particular, {\it tethered surfaces}, which model polymerized flexible
membranes, have unusual and interesting elastic properties.
While these properties are now well understood theoretically for
``phantom membranes", that is when self-avoidance (SA) interactions are ignored,
the consequence of incorporating SA constraints to describe
real membranes is still an open problem.
In practice, the search for a consistent theoretical treatment of SA
interactions raises the fundamental question of applying renormalization group
(RG) methods to {\it extended} objects, which is the issue addressed here.

The theoretical study of SA polymerized membranes is centered around
a model of tethered self-avoiding manifolds (SAM)\nref\rKN{M. Kardar 
and D. R. Nelson, Phys. Rev. Lett. {\bf 58}
(1987) 1289, 2280 (E); Phys. Rev. {\bf A 38}
(1988) 966.}
\nref\rArLub{J. A. Aronowitz and T. C. Lubensky, Europhys. Lett. {\bf 4}
(1987) 395.}[\xref\rKN-\xref\rArLub] 
directly inspired by the Edwards model for polymers
\ref\rSirSam{S. F. Edwards, Proc. Phys. Soc. Lond. {\bf 85} (1965) 613.}.
The surfaces 
are generalized to intrinsically $D$-dimensional {\it manifolds},
representing $D$-dimensional connected networks, whose nodes, labeled
by internal continuous coordinates $x \in \RR^D$, are
embedded in external $d$-dimensional space with position 
vector $\rvec (x)\in \RR^d$.
The associated continuum Hamiltonian $\CH$ is
\eqn\eEdwards{
{\raise.2ex\hbox{$\CH$}/\raise -.2ex\hbox{${\rm k}_{\rm B}T$}}
\ =\ {1\over 2}\,\int d^Dx\,\big(\nabla_x\rvec (x)\big)^2+
{b\over 2}\int d^Dx\int d^Dx'\ \delta^{d}\big(\rvec (x)-\rvec (x')\big)
\ ,}
with an elastic Gaussian term and a self-avoidance 
two-body $\delta$-potential with excluded volume parameter $b>0$, non-local in
``manifold space" $\RR^D$.

A finite upper critical dimension (u.c.d.) $d^\star$ for the SA
interaction exists only for manifolds with a continuous internal
dimension $0<D<2$.
Phantom manifolds ($b=0$) are {\it crumpled} with a finite 
Hausdorff dimension $d_H=2D/(2-D)$, and $d^\star = 2d_H$.
In\nref\rBDbis{B. Duplantier, Phys. Rev. Lett. {\bf 58} (1987) 2733,
and in \rJerus .}
[\xref\rKN-\xref\rArLub,\xref\rBDbis]
an $\epsilon$-expansion about $d^\star$ was performed via a
{\it direct renormalization} (DR) method adapted from
polymer theory 
\ref\rdesCloiz{J. des Cloizeaux, J. Phys. France {\bf 42} (1981) 635.}.
But many issues remain unanswered:
The consistency of the DR method is proven only for $D=1$ by the famous mapping
of \eEdwards\  onto a (zero component) ${({\bf \Phi}^2)}^2(\rvec )$ field theory
in external $d$-dimensional space
\ref\rDeGe{P.G. de Gennes, Phys. Lett. {\bf A 38} (1972) 339.}.
When $D\ne 1$, model \eEdwards\ can no longer be mapped onto a local field
theory, and the validity of RG methods and of scaling laws has been justified
only at leading order through explicit partial resummations
\ref\rDupHwaKar{B. Duplantier, T. Hwa and M. Kardar, Phys. Rev. Lett. {\bf 64}
(1990) 2022.}.
The questions of a proper treatment for boundaries and of the value of the
configuration exponent $\gamma$ \rBDbis\ are also open.

\medskip
In this Letter, we introduce a flexible formalism that allows us to prove
the validity of the RG approach to self-avoiding manifolds,
as well as to a larger class of manifold models with non-local interactions.
It broadly extends a recent work by the authors
\ref\rIIM{F. David. B. Duplantier and E. Guitter, Phys. Rev. Lett. {\bf 70}
(1993) 2205; Nucl. Phys. {\bf B394} (1993) 555.}
for a simpler model
\ref\rBD{B. Duplantier, Phys. Rev. Lett. {\bf 62} (1989) 2337.},
with a {\it local} singular interaction,
of a phantom manifold interacting with a single impurity
\ref\rLasLip{See also M. L\"assig and R. Lipowsky, Phys. Rev. Lett. {\bf 70}
(1993) 1131.}.
The present formalism is based on a new operator product expansion involving
{\it multilocal singular operators}, and allows for a systematic analysis of
the short distance ultraviolet (UV) singularities of the model.
At the critical dimension $d^\star$, we can classify all the relevant operators
and show that the model \eEdwards\ is {\it renormalizable to all orders} by 
renormalizations (i) of the coupling $b$
and (ii) of the position field $\rvec$.
As a consequence, we establish the validity of scaling laws for {\it infinite}
membranes, as well as the existence of finite size scaling laws for
{\it finite} membranes.
The latter result ensures the consistency of the DR approach.
A surprising result, which distinguishes manifolds with non-integer $D$ from
open linear polymers, is the absence of {\it boundary} operator
renormalization, leading to the general {\it hyperscaling relation}
\eqn\eHyper{\gamma=1-\nu d\, ,}
valid for finite SAM with $D<2$, $D\ne 1$.
Another surprise when considering SAM at the $\Theta$-point is
the appearance of a new relevant interaction term, which can supersede
the usual three-body term.

\medskip
\noindent{\bf Perturbation Theory.}
For infinite SA manifolds, physical observables are expressible in terms of the
$P$-point correlation functions, whose perturbative expansions are formally
\eqn\eCorrFirst{
\bigl\langle \sprod_{l=1}^{P}{\rm e}^{\ii \qvec_l\rvec(z_l)}\bigr\rangle
\ =\ {1\over \CZ}\,\sum_{N=0}^\infty\,{(-b)^N\over 2^N N!}\,
\int\sprod_{i=1}^{2N}{d^Dx_i}
\bigl\langle
\sprod_{l=1}^{P}{\rm e}^{\ii \qvec_l \rvec(z_l)}
\sprod_{a=1}^{N}\delta^{d}(\rvec(x_{2a})-\rvec(x_{2a-1}))
\bigr\rangle_0
}
The r.h.s. average $\langle\ldots\rangle_0$ is performed with respect to the
ideal Gaussian manifold ($b=0$).
The partition function $\CZ$ in the denominator has a similar perturbative
expansion in $b$, but with no external points.
The product of $\delta$ functions in \eCorrFirst\ 
can be written in terms of exponential operators as
\eqn\eExpRep{
\sprod_{a=1}^{N}\delta^{d}(\rvec(x_{2a})-\rvec(x_{2a-1}))=
\int\sprod_{i=1}^{2N}{d^d\kvec_i\over (2\pi)^d}
\sprod_{a=1}^{N}\CC_a\{\kvec_i \}
\sprod_{i=1}^{2N}{\rm e}^{\ii \kvec_i \rvec(x_i)}
}
with $N$ ``dipolar constraints"
$\CC_a\{\kvec_i\}=(2\pi)^d\delta^d(\kvec_{2a-1}+\kvec_{2a})$ for momenta
$\kvec_i \in \RR^d$ (later called ``charges") assigned to the points $x_i$.
The correlation function \eCorrFirst\ 
is defined as translationally invariant in external space, i.e.
with the ``neutrality rule" $\ssum\limits_{l=1}^{P}\qvec_l=\zerovec $,
a condition which is necessary when dealing with infinite membranes
to avoid infrared (IR) singularities.
The Gaussian average in \eCorrFirst\ is easily performed, using the identity
\eqn\eGausAv{
\bigl\langle \sprod\limits_{i}{\rm e}^{\ii\kvec_i\rvec(x_i)} \bigr\rangle_0 =
\exp\Big({-{1\over 2}\ssum\limits_{i,j}\kvec_i\kvec_j G_{ij}}\Big)
}
where
${G}_{ij}=-|x_i-x_j|^{2-D}/\big((2-D)S_{D}\big)$
is the massless propagator (Coulomb potential in $D$ dimensions),
with $S_{D}={2\,\pi^{D/2}\over\Gamma({{\scriptstyle D/2}})}$.
Integration over the momenta $\kvec_i$ then gives for the $N$'th
term of \eCorrFirst\ the manifold integral
\eqn\eManInt{
\int \sprod_{i=1}^{2N} d^D x_i\,\Delta^{-{d\over 2}}\,
\exp\left(-\,{1\over 2}\ssum\limits_{l,m=1}^{P}
\qvec_l\qvec_m\, {\Delta_{lm}\over\Delta}\right)
}
where $\Delta\{x_i\}$ is the determinant associated with the quadratic form 
(now on $\RR$) $Q\{k_i\}=\ssum\limits_{i,j=1}^{2N} k_i k_j{G}_{ij}$ 
restricted to the vector space defined by the $N$ neutrality constraints
$\CC_a\{k_i\}$, $k_{2a}+k_{2a-1}=0$,
and $\Delta_{lm}$ is a similar determinant involving also the
external points $z_l$ and $z_m$ \rDupHwaKar .

Note that a proper analytic continuation in $D$ of \eManInt\ 
is insured from \rIIM\ by the use of distance geometry,
where the Euclidean measure over the $x_i$ is understood as
the corresponding measure over the mutual squared distances
$a_{ij}=|x_i-x_j|^2$, a distribution analytic in $D$.

\medskip\noindent{\bf Singular Configurations.}
The integrand in \eManInt\ is singular when the determinant
$\Delta\{x_i\}\le 0$.
The associated quadratic form $Q\{k_i\}$,
restricted by the neutrality constraints $\CC_a\{k_i\}$, is the electrostatic
energy of a gas of charges $k_i$ located at $x_i$, and constrained to
form $N$ neutral pairs $a$ of charges (dipoles).
For such a globally neutral gas, the Coulomb energy is
minimal when the charge density is zero everywhere, i.e. when
the non zero charges $k_i$ aggregate into neutral ``atoms".
When $0<D<2$, the corresponding minimal energy is furthermore {\it zero},
which implies that the quadratic form $Q$ is non-negative and thus 
$\Delta\ge 0$.
Singular $\{x_i\}$ configurations, with $\Delta=0$, still exist when $Q$
is degenerate, which happens when some dipoles are assembled in 
such a way that, with appropriate non-zero charges, they still can build
neutral atoms.
This requires some of the points $x_i$ to coincide {\it and} the
corresponding dipoles to form at least one closed loop (Fig. 1).
This ensures that the only sources of divergences are
{\it short distance singularities}, and extends the Schoenberg theorem
used in \rIIM .
\def\legend{
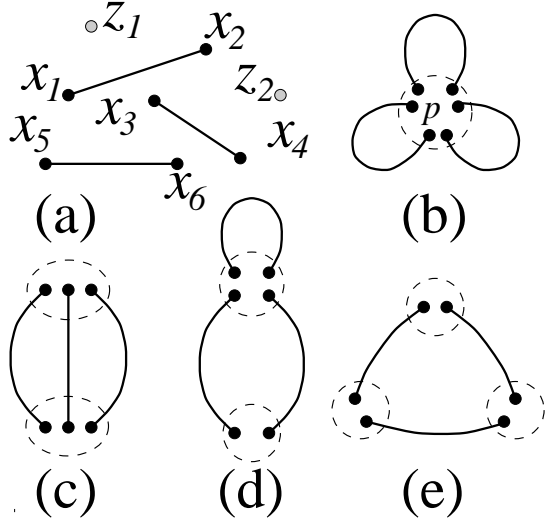
A general diagram with two external points and three internal dipoles (a);
``molecules" describing singular configurations with one (b), two (c,d)
and three (e) ``atoms".
(b,c,d) give UV divergences, (e) does not.}
\midinsert
\medskip
\centerline{\epsfbox{Fig.eps}}
\medskip
\noindent
{\bf Fig. 1:\ }
{\sl \legend}
\endinsert
\nfig\fOne{\legend}
\medskip\noindent{\bf Multilocal Operator Product Expansion.}
A singular configuration can thus be viewed as a connected ``molecule",
characterized by a set $\CM$ of ``atoms" $p$ with assigned positions $x_p$,
and by a set $\CL$ of links $a$ between these atoms, representing the
dipolar constraints $\CC_a$.
For each $p$, we denote by $\CP_p$ the set of charges $i$, at $x_i$, which
build the atom $p$ and define $y_i=x_i-x_p$ for $i\in \CP_p$.
The short distance singularity of $\Delta^{-d/2}$ is analyzed by performing a
small $y_i$ expansion of the product of the bilocal operators
$\varphi(x,x')\equiv \delta^d(\rvec (x)- \rvec (x'))$
for the links $a\in \CL$, in the Gaussian manifold theory (Eq. \eCorrFirst).
As will be shown below, this expansion around $\CM$ can be written as a 
{\it multilocal operator product expansion} (MOPE)
\eqn\eOE{\sprod_{a\in \CL}\varphi(x_{2a},x_{2a-1})=\sum_\Phi \Phi\{x_p\}
\,C^\Phi_{{\underbrace{\scriptstyle \varphi\dots\varphi}\atop |\CL|}}\{y_i\}}
where the sum runs over all multilocal operators $\Phi$ of the form:
\eqn\eMult{\Phi\{x_p\}=\int d^d\rvec \,
\sprod_{p\in\CM}\Big\{ \,
\npr \left\{\left({\bf \nabla_\rvec}\right)^{q_p}\delta^d(\rvec-\rvec (x_p))
\right\}\,A_p (x_p)\npr\Big\}
}
Here $A_p (x_p)\equiv A^{(r_p,s_p)}(\nabla_x,\rvec (x_p))$ is a local operator at
point $x_p$, which is a product of $x$-derivatives of the field $\rvec$, 
of degree $s_p$ in $\rvec (x_p)$ and degree $r_p\ge s_p$ in $\nabla_x$.
$({\bf \nabla_\rvec})^{q_p}$ denotes a product of $q_p$ derivatives with
respect to $\rvec$, acting on $\delta^d(\rvec -\rvec (x_p))$. The symbol
``$\npr\quad\npr$" denotes the {\it normal product} subtraction prescription at $x_p$
(which, in a Gaussian average, amounts to setting to zero any derivative of 
the propagator ${G}_{ij}$ at coinciding points $x_i=x_j=x_p$).
For ${\rm Card}(\CM)\equiv|\CM|>1$, \eMult\ describes the most general
$|\CM|$-body contact
interaction between the points $x_p$, with possible inserted local operators 
$A_p (x_p)$ at each point $x_p$.
For $|\CM|=1$, it reduces to a local operator $A_p(x_p)$.

The coefficient associated with the operator $\Phi$ in the MOPE,
$C^\Phi_{\varphi\dots\varphi}\{y_i\}$, 
can be written as an integral over the momenta $\kvec_i$:
\eqn\eCoeff{ C^{\Phi}_{\varphi\dots\varphi}\{y_i\} =
\int\kern -.5em 
\sprodp_{a\in\CL}
\CC_a\{\kvec_i\}\kern -.2em
\sprod_{p\in\CM}\Bigg\{\sprod_{i\in\CP_p}\kern -.2em d^d\kvec_i
\big\{({\bf \nabla_{\kvec}})^{q_p}
\delta^d(\ssum_{i\in\CP_p}\kvec_i)\big\}
C^{A_p}\{y_i,\kvec_i\}\,
{\rm e}^{-{1\over 2}{\kern -1.5em} \sssum\limits_{\ \quad i,j\in\CP_p}
{\kern -1.5em} \kvec_i\kvec_j {G}_{ij}}\Bigg\}
}
where $C^{A_p}\{y_i,\kvec_i\}$
is a monomial associated with the operator $A_p$, of similar
global degree $r_p$ in the $y_i$, and $s_p$ in the $\kvec_i$.
The product $\sprodp$ is over all constraints $a\in \CL$ but one.

The MOPE \eOE\  follows from the expression \eExpRep\ 
in terms of free field exponentials plus constraints.
For each $p$, we use the general small $y_i$ local operator product identity
\eqn\eOPEVO{
\sprod_{i\in \CP_p}{\rm e}^{{\rm i}\,\kvec_i\rvec(x_i)} \ =\ 
{\npr}\left.\sprod_{i\in \CP_p}
{\rm e}^{\left(y_i{\partial\over\partial x_i}\right)}
{\rm e}^{{\rm i}\,\kvec_i\rvec(x_i)}
\right|_{x_i=x_p}
{\hskip -1.5em\npr}\ 
{\rm e}^{-{1\over 2}{\kern -1.5em} \sssum\limits_{\ \quad i,j\in\CP_p}
{\kern -1.5em} \kvec_i\kvec_j G(y_i,y_j)}
}
When expanded in the $y_i$, the normal product $\npr (\ )|_{x_i=x_p}\npr$
in \eOPEVO\ gives a sum
$\ssum\limits_{A} C^A\{y_i,\kvec_i\}\break
\npr A(x_p) {\rm e}^{\ii\kvec_p\rvec_p}\npr$
(denoting $\kvec_p=\ssum\limits_{i\in\CP_p}\kvec_i$ and
$\rvec_p\equiv\rvec(x_p)$)
which generates the local operators $A(x_p)$ and the monomials $C^A$ of
\eMult\ and \eCoeff .
We insert the identity
$1\equiv\int d^d\,\kvec_p \delta^d(\kvec_p-\ssum\limits_{i\in \CP_p} \kvec_i)$
in \eExpRep\ for each atom $p\in\CM$,
rewrite one of the dipolar constraints as a global neutrality constraint
$\delta^d(\ssum\limits_{p\in\CM}\kvec_p)$
on the $\kvec_p$,
and expand each
$\delta^d(\kvec_p-\ssum\limits_{i\in\CP_p}\kvec_i)$ in powers of $\kvec_p$.
Finally by integrating over the $\kvec_p$, the constraint
$\delta^d(\ssum\limits_{p\in\CM}\kvec_p)$
builds the multilocal $|\CM|$-body operator $\Phi\{x_p\}$ and we obtain the MOPE
\eOE , \eMult\ and \eCoeff .

\medskip\noindent{\bf Power Counting and Renormalization.}
The MOPE \eOE\ allows us to determine those singular configurations which give
rise to actual UV divergences in the manifold integral \eManInt .
Indeed, given a singular configuration $\CM$ and integrating
over the domain where the relative positions $y_i=x_i-x_p$ are of order
$|y_i|\lesssim\rho$, we can use the MOPE
of \eExpRep\ to obtain an expansion of the integrand in \eCorrFirst\ 
in powers of $\rho$.
Each coefficient $C^\Phi_{\varphi\dots\varphi}$
gives a contribution of order 
$\rho^{\omega}$, with degree $\omega$ given by power counting
\eqn\eDegree{\omega=D\{2|\CL|-|\CM|\}+d\nu_0\{|\CM|-|\CL|-1\}+
\ssum_{p\in \CM}\big\{\nu_0 (q_p-s_p)+r_p\big\}}
with $\nu_0\equiv(2-D)/2<1$ and $r_p\ge s_p$.
Whenever $\omega\le 0$, a UV divergence occurs, as a factor multiplying the
insertion of the corresponding operator $\Phi$.
At the upper critical dimension $d^\star=2D/\nu_0$, $\omega$ becomes
independent of the number $|\CL|$ of dipoles, and is equal to the canonical
dimension $\omega_\Phi$ of $\int\!\sprod\limits_{\CM}\!d^Dx\,\Phi$
in the Gaussian theory.
Three relevant operators, with
$\omega_\Phi\le 0$ and such that the corresponding coefficient does 
not vanish by symmetry, are found by simple inspection.
Two of these operators are {\it marginal} ($\omega_\Phi=0$):
(i) the two-body SA interaction term $\delta^d(\rvec_p-\rvec_{p'})$ itself, 
obtained through singular configurations with $|\CM|=2$ atoms (and
with $q\!=\!r\!=\!s\!=\!0$ for $p$ and $p'$),
(ii) the one-body local elastic term 
$\npr (\nabla \rvec_p)^2 \npr$, obtained for $|\CM|=1$ ($q\!=\!0$,
$r\!=\!s\!=2$).
The third operator is {\it relevant} with $\omega_\Phi=-D$, and is just the
identity operator {\bf 1} obtained when $|\CM|=1$ ($q\!=\!r\!=\!s\!=\!0$).
It gives ``free energy" divergences proportional to the manifold volume,
which cancel out in IR finite observables \eCorrFirst .

The above analysis deals with {\it superficial UV divergences} only.
A complete analysis of the general UV singularities associated with
successive contractions toward ``nested" singular configurations
can be performed, using the techniques of \rIIM\ and the fact that an 
iteration of the MOPE only generates multilocal operators of the type \eMult .
The results are:
(i) that the observables \eCorrFirst\ are UV finite for $d<d^\star(D)$, and
are meromorphic functions in $d$ with poles at $d=d^\star$,
(ii) that a renormalization operation, similar to the subtraction operation
of \rIIM , can be achieved to remove these poles, (iii) that this operation
amounts to a renormalization of the Hamiltonian \eEdwards .
More explicitly, the renormalized correlation functions
$\blangle\sprod\limits_{l=1}^P{\rm e}^{\ii \qvec_l\rvecR(z_l)}\brangle_{\bf R}$
have a finite perturbative expansion in the renormalized coupling $\bR$,
when $\blangle\cdots\brangle_{\bf R}$ is the average w.r.t the renormalized
Hamiltonian
\eqn\eRenHam{
{\raise.2ex\hbox{$\CH_{\bf R}$}/\raise -.2ex\hbox{${\rm k}_{\rm B}T$}}
\ =\ {Z\over 2}\,\int d^Dx\,\big(\nabla_x\rvecR (x)\big)^2+
{1\over 2}\,{\bR\mu^\epsilon Z_b}
\int d^Dx\int d^Dx'\ \delta^{d}\big(\rvecR (x)-\rvecR (x')\big)
\ .}
$\mu$ is a renormalization (internal) momentum scale, $\epsilon=2D-d\nu_0$,
$Z_b(\bR)$ and $Z(\bR)$ are respectively the coupling constant and
the field renormalization factors, singular at $\epsilon=0$.
At first order, we find by explicitly calculating
$C^{(\nabla\rvec)^2}_{\varphi}$ and $C^\varphi_{\varphi\varphi}$ that
$Z=1+\bR{B \over \epsilon}{(2-D)^2\over 2D}$,
$Z_b=1+\bR{B \over \epsilon}{\Gamma^2(D/(2-D))\over \Gamma(2D/(2-D))}$,
with $B={1\over 2}(4\pi)^{-{d\over 2}}S_D^{2+{d\over 2}}(2-D)^{-1+{d\over 2}}$.
For the quantities which are not IR finite, which we discuss later, an
additive counterterm proportional to the volume of the manifold
(corresponding to the relevant identity operator {\bf 1})
is also necessary.

Expressing the observables of the SAM model \eEdwards\ in terms of
renormalized variables $\rvec=Z^{1/2}\rvecR$, $b=\bR\mu^\epsilon Z_b Z^{d/2}$,
one can derive in the standard way RG equations involving Wilson's functions
$W(\bR)=\raise.2ex\hbox{${\scriptstyle\mu}$}
{\partial\over\partial\mu}\bR\big|_{b}$,
$\nu(\bR)=\nu_0-{1\over 2}\raise.2ex\hbox{${\scriptstyle\mu}$}
{\partial\over\partial\mu}\ln Z\big|_{b}$.
A non-trivial IR fixed point $\bR^\star\!\propto\!\epsilon$ is found for
$\epsilon>0$.
It governs the large distance behavior of the SA infinite manifold, which
obeys scaling laws characterized by the exponent
$\nu$, defined for instance through the 2-point function
$\blangle (\rvec(x)-\rvec(0))^2\brangle\propto |x|^{2\nu}$.
The value obtained in this approach,
$\nu=\nu(\bR^\star)$, corroborates that obtained in
[\xref\rKN-\xref\rArLub,\xref\rDupHwaKar] at first order in $\epsilon$.

\medskip\noindent{\bf Finite Size Scaling and Direct Renormalization.}
The DR formalism requires one to consider {\it finite} manifolds with ``internal
volume" $V$, and to express scaling functions in terms of a dimensionless
second virial coefficient $g=-R_G^{-d}\CZ_{2,c}/(\CZ_1)^2$,
where $\CZ_1(V)$ and $\CZ_{2,c}(V)$ are respectively the one- and two
membrane (connected) partition functions, and $R_G$ is the radius of gyration.

\def\texFoot{
The expansion at the origin of the massless propagator ${\tilde G}$ on a curved
manifold reads in Riemann normal coordinates
${\tilde G}(x)\simeq G(x)-{|x|^2\over 2D}\blangle\npr(\nabla{\bf r})^2\npr
\brangle$,
with $G(x)\propto |x|^{2-D}$ the propagator in infinite flat space, and
next order terms $\CO(|x|^{4-D})$ proportional to the curvature and
subdominant for $D<2$; the normal product $\npr\quad\npr$ is still defined
w.r.t. infinite flat space,
and gives explicitly for a finite manifold with volume $V$ 
$\blangle\npr(\nabla{\bf r})^2\npr\brangle=-1/V$.
}
When dealing with a finite closed manifold (for instance the $D$-dimensional
sphere $\CS_D$ \rIIM), characterized by its (in general curved) internal metric,
the massless propagator $G_{ij}$ gets modified.
Nevertheless, from \eOPEVO\ and the short distance expansion of
$G_{ij}$ in a general metric\ref\rFootPop{\texFoot},
one can show that
the short distance MOPE \eOE\ remains valid, provided that
the sum is extended to include multilocal operators $\Phi$ still of the form
\eMult , but with local operators $A(x)$ involving also the Riemann curvature
tensor and its derivatives, with appropriate coefficients
$C^\Phi_{\varphi\ldots\varphi}$.
A crucial point is that in the MOPE the dependence on the geometry of the
manifold (size, curvature,$\ldots$) is encoded only in the expectation values
$\blangle\ldots\Phi\ldots\brangle_0$ of the multilocal operators $\Phi$,
while the short distance behavior ($y_i\to 0$) of coefficients
$C^\Phi_{\varphi\ldots\varphi}\{y_i\}$ is {\it independent} of the geometry.
Thus, at $d^\star$, UV divergences still come with insertions of
relevant multilocal operators with $\omega_\Phi\le 0$.

When $0<D<2$, none of the new operators involving the
curvature is found to be relevant by power counting.
Therefore, the infinite membrane counterterms $Z$ and $Z_b$ still
renormalize the finite membrane theory.
Since, as for finite size scaling
\ref\rFSS{E. Br\'ezin, J. Physique {\bf 43} (1982) 15.}, the manifold size is
not renormalized, arguments parallel to those of
\ref\rBenMah{M. Benhamou and G. Mahoux, J. Physique {\bf 47} (1986) 559.}
for polymers can be used to justify the DR formalism.
Indeed, the second virial coefficient $g(b,V)$ (as well as any
{\it dimensionless} scaling function) must be UV finite when expressed as a
function $g_{\hbox{\fivebf R}}(\bR, V\mu^D)$
of the renormalized coupling $\bR$ (and of $\mu$).
As a consequence, (i) the scaling functions are finite when expressed in terms
of $g$ and obey RG equations.
The existence of a non-trivial IR fixed point $\bR^\star$ for
$\epsilon>0$ implies that (ii) in the large volume limit $V\to\infty$, $g$ tends
toward a finite limit $g^\star=g_{\hbox{\fivebf R}}(\bR^\star)$
(independent of $V\mu^D$), and so do all scaling functions.
Points (i) and (ii) are the essence of DR.

\medskip\noindent{\bf Hyperscaling.}
As mentioned above, the renormalization of partition functions for a
finite SAM requires an additional counterterm (shift of the free
energy) proportional to the manifold volume $V$. 
A consequence of the absence of other geometry dependent relevant operators
when $0<D<2$ is the general hyperscaling law \eHyper\ valid for {\it closed}
SAM, and relating the configuration exponent $\gamma$,
defined by
\def\sumconf{
\kern -3em \sum_{\quad\qquad\hbox{\sevenrm Configurations}}\kern -3em}
\eqn\eDefGam{\CZ_1(V)=\int\CD[\rvec]\,
\delta^d(\rvec(0))\, {\rm e}^{-\CH/k_BT}\sim V^{{\gamma-1\over D}}\ ,
}
to the exponent $\nu$.
Indeed, from \eDefGam , once the free energy divergent term has been
subtracted, $\CZ_1$ is simply multiplicatively renormalized as
$\CZ_1(b,V)=Z^{-d/2}\CZ_1^{\bf R}(\bR,V\mu^D)$.
This validates the scaling hypothesis that
$\CZ_1\sim |\rvec|^{-d}\sim V^{-\nu d/D}$,
and leads directly to \eHyper .

For {\it open} SAM with {\it free} boundaries, and when $1\le D<2$, the
boundary operator
$\int_{\raise-.3ex \hbox{\fiverm boundary}}\kern-2.5em d^{D-1}x\,\hbox{\bf 1}$
becomes relevant.
Since it is simply a
geometrical quantity, it cannot modify the renormalizations of $\rvec$ and $b$.
Furthermore, it is marginally relevant only for $D=1$ \rBDbis\ and therefore, 
as long as $D\ne 1$ the scaling laws and the hyperscaling relation 
\eHyper\ {\it remain valid}.
Only at $D=1$, the corresponding end-point counterterm enters the multiplicative
renormalization of $\CZ_1$, and $\gamma$ becomes an independent exponent,
with an extra contribution from the two end-points.

Eq. \eHyper\ has been checked explicitly at order $\epsilon$ for the sphere
$\CS_D$ and the torus $\CT_D$.
Previous calculations [\xref\rKN,\xref\rArLub],
which yield \rBDbis\ $\gamma=1$ for non-integer $D$, did not involve
the physical massless propagator\rFootPop\ (valid for a finite manifold with
Neumann boundary conditions), used here.

When $D\ge 2$, if the small $\epsilon$ RG picture remains valid, i.e. if the
large distance properties of SAM are governed by the IR fixed point
$\bR^\star$, operators involving curvature become relevant, and 
\eHyper\ is not expected to be valid, even for closed manifolds.
\medskip
\noindent{\bf $\Theta$-point and long-range interactions.}
The above formalism is directly applicable to a large class of manifold models
where the interaction can be expressed in term of free field exponentials
with suitable neutrality constraints $\CC_a\{\kvec_i\}$. 
Examples of such interactions are the $n$-body contact potentials 
but also the two-body 
long-range Coulomb potential $1/|\rvec-\rvec '|^{d-2}$, which can be
represented by modified dipolar constraints 
$\CC\{\kvec_i\}=|\kvec|^{-2}\delta^d(\kvec+\kvec')$.
For all these models, the MOPE involves the same multilocal operators as in
\eMult , with modified coefficients (still given by \eCoeff , but with new
constraints $\CC_a$).

As an application, we may ask for the most relevant short-range
interaction describing a polymerized membrane at the $\Theta$-point, i.e.
when the two-body term $b$ in \eEdwards\ vanishes.
It is either the usual three-body contact potential, with u.c.d.
$d_3^\star=3D/(2-D)$,
as for ordinary polymers, or the two-body singular potential 
\hbox{$\Delta_{\rvec}\delta^d(\rvec-\rvec ')$} with u.c.d.
${\tilde d}^\star_2=2(3D-2)/(2-D)$, which indeed is the most relevant one when
$D>4/3$.

Finally, the absence of long-range potentials in the MOPE shows that long-range
interactions are not renormalized.
For instance, when considering charged polymerized membranes with a two-body
Coulomb potential, the only (marginally) relevant operator at the u.c.d.
is the local operator $\npr (\nabla \rvec )^2 \npr$, indicating that only
$\rvec$ is renormalized.
As a consequence, it is easy to show that $\nu=2D/(d-2)$ exactly in this case.

\medskip\noindent{\bf Acknowledgements:}

We thank J. Miller for his interest and a careful reading of the manuscript.

\vfill\eject
\listrefs
\bye